\documentclass[aps,
english,preprintnumbers,nofootinbib,
twocolumn]{revtex4-1}

\usepackage{amsfonts,amsmath,amssymb}
\usepackage{graphicx}
\usepackage[utf8]{inputenc}
\usepackage{hyperref}
\usepackage{babel}
\usepackage{empheq}

\usepackage{xcolor}

\newcommand\be{\begin{equation}}
\newcommand\ee{\end{equation}}
\newcommand\bea{\begin{eqnarray}}
\newcommand\eea{\end{eqnarray}}

\begin{document}

\title{ Krylov Complexity as a Probe for Chaos}

\author{Mohsen Alishahiha${}^a$,
Souvik Banerjee${}^{b,c}$
and,  Mohammad Javad Vasli${}^{d,a}$
}
\affiliation{${}^a$ School of Physics, Institute for Research in Fundamental Sciences (IPM),\\
	P.O. Box 19395-5531, Tehran, Iran\\ 
 ${}^b$ Institut f{\"u}r Theoretische Physik und Astrophysik,
	Julius-Maximilians-Universit{\"a}t W{\"u}rzburg,\\ Am Hubland, 97074 W{\"u}rzburg, Germany\\ ${}^c$ W\"urzburg-Dresden Cluster of Excellence ct.qmat\\
 ${}^d$  Department of Physics, University of Guilan, P.O. Box 41335-1914, Rasht, Iran\\
 E-mails: {\tt  alishah@ipm.ir,
 souvik.banerjee@uni-wuerzburg.de, vasli@phd.guilan.ac.ir
}}

\begin{abstract}
In this work, we explore in detail, the time evolution of Krylov complexity. We demonstrate, through analytical computations, that in finite many-body systems, while ramp and plateau are two generic features of Krylov complexity, the manner in which complexity saturates reveals the chaotic nature of the system. In particular, we show that the dynamics towards saturation precisely distinguish between  chaotic and integrable systems. For chaotic models, the saturation value of complexity reaches its infinite time average at a finite saturation time. In this case, depending on the initial state, it may also exhibit a peak before saturation. In contrast, in integrable models, complexity approaches the infinite time average value from below at a much longer timescale. We confirm this distinction using numerical results for specific spin models.

\end{abstract}

\maketitle

\section{Introduction}
\label{sec:intro}

While we have a solid understanding of classical 
chaos, quantum chaos remains a more elusive concept. 
Classical chaotic behaviour is characterized by the 
sensitivity of trajectories in phase space to initial 
conditions, where two trajectories initially close to each other
diverge exponentially fast — a phenomenon quantified by 
the Lyapunov exponent. In contrast, quantum chaos is a 
more ambiguous concept and is rather 
challenging to understand. This is primarily due to the absence of a 
clear understanding of quantum phase space.

In an isolated quantum system defined by its Hamiltonian, along with specific boundary or initial conditions, it is natural to expect that the key indicators of its chaotic nature are encoded in the structure of its eigenstates and eigenvalues. Indeed, it is well known that
the eigenvalues of chaotic Hamiltonians display distinct statistical features. More precisely,
the level spacing of a maximally chaotic quantum system follows the Wigner-Dyson distribution, while that of integrable systems follows the Poisson distribution
\cite{Bohigas:1983er}.

To investigate the nature of quantum chaos, several 
quantities have been introduced, including out-of-time-
order correlators (OTOCs) 
\cite{Shenker:2013yza,Shenker:2013pqa}. 
Semiclassically, OTOCs exhibit exponential growth due 
to the butterfly effect, characterized by the Lyapunov 
exponent, which is conjectured to be bounded 
\cite{Maldacena:2015waa}. This bound is saturated in 
certain strongly interacting models with holographic 
descriptions, such as the Sachdev-Ye-Kitaev (SYK) 
model \cite{Kitaev,Sachdev:1992fk,Maldacena:2016hyu}. 
However, it is important to note that the exponential 
growth of OTOCs is not a universal characteristic of 
all chaotic systems
\cite{Fine_2014,Xu_2019,Hashimoto:2020xfr}.

The present article aims to explore the
chaotic nature of 
many-body quantum systems using Krylov 
complexity\footnote{Throughout this paper, we will 
	examine the Krylov complexity of states 
	\cite{Balasubramanian:2022tpr,Alishahiha:2022anw,Caputa:2024vrn}. }, 
as defined in the context of the Lanczos algorithm. While 
this method has traditionally been employed for 
numerical computations \cite{viswanath1994recursion}, 
in recent years there has been increased interest 
in applying the Krylov method to the study of quantum chaos  \cite{Parker:2018yvk} (for review see \cite{Nandy:2024htc}) \footnote{For applications of Krylov method to entanglement generation and quantum metrology
	see {\it e.g.} \cite{Chu:2024cah}.}. 

The authors of \cite{Parker:2018yvk} proposed a universal operator growth hypothesis that connects the asymptotic behaviour of Lanczos coefficients to the dynamics of the system in question. Specifically, for chaotic many-body quantum systems in dimensions greater than one and lacking symmetry, the Lanczos coefficients are expected to exhibit asymptotic linear growth. However, while the literature supports this linear growth in chaotic systems, it may not be a universal indicator of chaos; non-chaotic models can also display asymptotic linear
growth \cite{Dymarsky:2021bjq,Bhattacharjee_2022,
	Avdoshkin:2022xuw,Camargo:2022rnt,Vasli:2023syq}.

The Krylov complexity saturates in  many-body systems 
when the Lanczos coefficients vanish at certain steps. 
It is therefore natural to examine this saturation 
value to determine whether it can distinguish between 
chaotic and integrable models. Actually, 
in an explicit 
Ising model, it has been demonstrated that the 
saturation value of the Krylov complexity increases as 
one transitions from an integrable model to a chaotic 
one \cite{Rabinovici:2021qqt,Rabinovici:2022beu}. 
However, it is important to note that the saturation 
value of Krylov complexity also depends on the 
specific state for which it is computed. In fact, the 
saturation value for chaotic systems may or may not 
exceed that of integrable models 
\cite{Rabinovici:2021qqt,Rabinovici:2022beu,
	Scialchi:2023bmw,Ballar_Trigueros_2022,Espa_ol_2023}.
We will return to this point later.

Using the evolution of maximally entangled states in 
the Krylov basis, it was shown in 
\cite{Erdmenger:2023wjg} that the linear growth and 
saturation of Krylov complexity are generic features 
of many-body systems, regardless of whether they are 
chaotic or integrable. However, in chaotic systems, 
the Krylov complexity exhibits a peak before reaching 
its saturation value at late times 
\cite{Balasubramanian:2022tpr}.

The presence of this peak before saturation has 
recently been observed numerically in various  
models 
\cite{Huh:2023jxt,Camargo:2024deu,Nandy:2024wwv,
	Bhattacharjee:2024yxj,Balasubramanian:2024ghv,
	Baggioli:2024wbz}. It is important to note that the appearance of a peak prior to saturation is not solely 
determined by the system's dynamics; it also depends 
on the initial state. For instance, it occurs for the initial state considered in \cite{Erdmenger:2023wjg}. Nonetheless, it appears that the 
thermofield double state at infinite temperature 
provides a maximal universal bound on this peak
\cite{Camargo:2024deu}.

In this paper, we aim to further explore how Krylov 
complexity can probe the chaotic nature of systems. To 
achieve this, we leverage the fact that Krylov 
complexity is essentially the expectation value of the 
number operator in the Krylov basis. Consequently, 
studying the time dependence of complexity reduces to 
analyzing the time evolution of this expectation 
value—an object we are more familiar with in the context of quantum
thermalization
(see {\it e.g.} \cite{Alishahiha:2022nhe}).

Inspired by the procedure that leads to 
the eigenstate thermalization hypothesis (ETH)
\cite{Srednicki:1994mfb,Deutsch:1991}, we find that 
the growth and subsequent saturation of complexity at 
a time exponentially large in relation to the system 
size is a generic feature of Krylov complexity, 
regardless of whether the dynamics is chaotic or 
integrable. The specific aspect of complexity that is 
sensitive to the chaotic nature of the system is how 
it approaches its saturation value at late times. In 
chaotic systems, the saturation value of complexity
reaches the infinite time average of the complexity  
at the saturation time, whereas in integrable 
models, it approaches this value from below over a 
relatively longer time.  

It is worth noting that while the aforementioned 
proposal outlines the generic features of complexity 
for chaotic and integrable models, the detailed 
behaviour of complexity also depends on the initial 
state. In the chaotic case, for a maximally entangled 
state, the Krylov complexity exhibits a clear peak 
before reaching its saturation value. However, for 
generic initial states, this peak may or may not 
exist, or it may not be pronounced. Additionally, 
there are small fluctuations around the saturation 
value, the shape and nature of which also depend on 
the initial state.

The rest of the paper is organized as follows.  In section \ref{Krylov:review}, we briefly review the Lanczos algorithm and present a closed analytic form for the expansion coefficient  
of Krylov basis in terms of the energy eigenstates, for a general closed many body system. This makes the perfect background on which in section \ref{sec:Krylov-chaos} we study several generic features of Krylov complexity leading to the main claim of this paper. In section \ref{example}
we examine our proposal
using numerical results from an explicit many-body
system. The last section is devoted to discussions.

\section{Krylov basis vs energy eigenbasis}
\label{Krylov:review}

To set the stage,  let us consider a  closed quantum system with a time-independent local 
non-degenerate Hamiltonian $H$ whose eigenstates and eigenvalues 
are denoted by $|E_\alpha\rangle$, and $E_\alpha$, respectively.
Here $\alpha=1,\cdots, {\cal D}$ with   ${\cal D}$ being  dimension of full Hilbert space.
Starting with an initial state, 
$|\psi_0\rangle$ at $t = 0$, its time evolution in the Schr\"odinger picture is given by
\be\label{Schr}
|\psi(t)\rangle=e^{-iHt}|\psi_0\rangle\,.
\ee 
At $t = 0$, the initial state can be expanded in terms of the energy eigenbasis as
\be
\label{eq:psi0-Ebasis}
|\psi_0\rangle=\sum_{\alpha=1}^{\cal D}c_\alpha|E_\alpha\rangle\,.
\ee

Starting with the same initial state $|\psi_0\rangle$ 
in the same quantum system with time-independent Hamiltonian $H$, the Krylov basis, $\{ |n\rangle, n=0,1,2,\cdots, 
{\cal D}_\psi-1\}$ can be constructed as follows. The first element of the basis is identified with the initial state $|0\rangle=|\psi_0\rangle$ which we assume to be normalized.
The other elements are then constructed, recursively,
following the algorithm \cite{viswanath1994recursion}
\be\label{GS}
(H-a_n)|n\rangle=b_{n+1}
| n+1\rangle +b_n|n-1\rangle\,,
\ee
where $a_n=\langle n|H|n\rangle$, and
\be\label{LC-state}
b^2_{n+1}=\langle n|H^2|n\rangle-a^2_n-b_n^2\,\;\;\;\;\;
{\rm with}\;b_0=0.
\ee
Note that, physically, $a_n$ is the average energy associated with the  state $|n\rangle$.
Moreover, denoting the variance of energy for this state as $\Delta a_n$ with
\be
(\Delta a_n)^2=\langle n|H^2|n\rangle-\langle n|H|n\rangle^2\,,
\ee
one gets
\be
b^2_{n+1}=\sum_{j=0}^{n}(-1)^j(\Delta a_{n-j})^2\,.
\ee

This recursive procedure stops whenever $b_n$ vanishes which occurs for $n={\cal D}_\psi\leq {\cal D}$ which defines the dimension of Krylov space.
Note that this algorithm produces an orthonormal and ordered basis together with coefficients $a_n$ and $b_n$ known as Lanczos coefficients \cite{Lanczos:1950zz}.

Once we have constructed the Krylov basis, the evolved state at any time can be expanded in this basis 
\be\label{phi}
|\psi(t)\rangle=\sum_{n=0}^{{\cal D}_\psi-1}\phi_n(t)\,|n\rangle\,,\;\;\;\;{\rm with }\;\;\sum_{n=0}^{{\cal D}_\psi-1}|\phi_n(t)|^2=1\,,
\ee
where the dynamics of the wave function $\phi_n(t)$ is governed by the Schr\"odinger equation
\be\label{Sch-phi}
-(i\partial_t+a_n)\phi_n(t)=b_{n}\phi_{n-1}(t)+b_{n+1}\phi_{n+1}(t)\,,
\ee
which should be solved using the initial  condition 
$\phi_n(0)=\delta_{n0}$.

An advantage of working with the Krylov basis lies in the fact that under a unitary time evolution, the trajectory of a given initial state does not necessarily spread into
the entire Hilbert space. Instead, it remains confined within a subset known as the Krylov space. The latter typically has a smaller dimension compared to the full Hilbert space of the system. Thus, following the argument above, the Krylov space suffices for studying the time evolution of the system. In particular, if the system has conserved charges, working in this basis we are 
automatically confined in a subsystem that preserves the symmetry of the initial state. 

To elaborate on the last point, let us expand an element of the Krylov space in terms of the energy eigenbasis
\be\label{NE}
|n\rangle=\sum_{\alpha=1}^{{\cal D}} f_{n\alpha} |E_\alpha\rangle\,.
\ee
Note that since ${\cal D}_\psi\leq {\cal D}$, the expansion coefficient, $f_{n\alpha}$, is not 
necessarily invertible and therefore, in general, energy eigenstates
cannot be expanded in terms of Krylov basis. Using the orthogonality condition 
of the Krylov basis one gets
\be
\sum_{\alpha=1}^{{\cal D}}f^*_{n\alpha}f_{m\alpha}=\delta_{nm}\,.
\ee

However, it is important to note that, on general grounds, the 
condition ${\cal D}_\psi \leq {\cal D}$ arises from a 
symmetry in the model and this can be understood as follows. Let us 
assume our model has a symmetry generated by the charge 
$Q$. Given this symmetry, we have $[Q, H]
=0$. This means that starting with an initial state of 
a specific symmetry charge, the entire Krylov basis is constrained to maintain that same charge.

This can be argued by noting that the energy eigenstates can actually be organized into 
different blocks according to their symmetry charges. 
Therefore, selecting a particular initial state with a 
definite charge, its spreading through the Krylov basis nevertheless retains 
that same charge. The dimension of the Krylov space coincides with that of the corresponding symmetry block \cite{PhysRevB.103.134207}. As an immediate consequence of this fact, the 
summation in the equation  \eqref{NE} terminates
at ${\cal D}_{\psi}$
\be\label{nE}
|n\rangle=\sum_{j=1}^{{\cal D}_\psi} f_{nj} |E_j\rangle\,,
\;\;\;\;\;\;\;{\rm for}\;\;n\leq {\cal D}_\psi-1.
\ee
In this case, $f_{nj}$ can be inverted within a fixed charge sector and 
accordingly, in this symmetry block, the energy eigenvectors can also
expanded in terms of the Krylov basis. Note that, within this sector, one still has the orthogonality relation,
\be\label{Orth}
\sum_{j=1}^{{\cal D}_\psi}f^*_{nj}f_{mj}=\delta_{nm}\,.
\ee

Following this justification, we will consider an initial state with a definite symmetry charge, which restricts the summation over energy eigenstates in \eqref{eq:psi0-Ebasis} to the fixed charge subsystem with dimension
${\cal D}_\psi$, the dimension of Krylov 
subspace, leading to an expansion
\be 
\label{eq:psi0-Dpsi}
|\psi_0\rangle =|0\rangle=\sum_{j=1}^{{\cal D}_\psi} c_{j} |E_j\rangle.
\ee

Plugging \eqref{nE} in \eqref{GS} yields\footnote{In our notation, $\{n,m\} = 0,\cdots {{\cal D}_\psi-1}$ while $\{i,j\} = 1,\cdots {{\cal D}_\psi}$. The Greek indices $\{\alpha,\beta\}$ are reserved for the full Hilbert space, namely, $\{\alpha,\beta \} = 1, \cdots {\cal D}$. }
\be\label{fnj}
(E_j-a_n)f_{nj}=b_{n+1}f_{n+1\,j}+b_n
f_{n-1\,j}\,,
\ee
for $n\leq {\cal D}_\psi-1$, that can be solved to find $f_{nj}$ in terms of $f_{0j}=
c_j$ as
\be\label{fnj1}
f_{nj}=\frac{g_n(E_j)}{b_n!}c_j\,.
\ee
Here we use the notation  that $b_n!=b_1\cdots b_n$ and $b_0!=1$, and $g_n(E_j)={\rm det}(E_j-H^{(n)})$, 
$H^{(n)}$ being an $n\times n$  matrix whose elements are given by
\be
H^{(n)}_{pq}=\langle p |H| q\rangle
=a_q 
\delta_{pq}+b_{q+1}\delta_{pq+1}+b_q\delta_{pq-1}\,,
\ee
for $p,q=0,\cdots,n-1$.  Substituting \eqref{fnj1} into \eqref{Orth} and \eqref{fnj}, one obtains an orthogonality relation for the function $g_{n}(E_i)$ and a recursion relation, respectively, of the forms\footnote{Equation \eqref{eq:continuant} can be immediately recognized as the recursion relation of the continuant of a tridiagonal matrix. This is consistent with the fact that Hamiltonian has a tridiagonal form in Krylov basis \cite{Balasubramanian:2022dnj}. We also observe that this recurrence relation 
	actually resembles the three-term recurrence relation for Krawtchouk polynomials 
	\cite{miki2022singleindexedexceptionalkrawtchoukpolynomials}. 
	This relationship can, in turn, be utilized to derive an exact 
	solution for $a_n$ and  $b_n$.}
\bea\label{SF}
&&\sum_{i=1}^{{\cal D}_\psi}|c_i|^2 g_{n}(E_i)g_{m}(E_i)=\left(b_n!\right)^2 \,\delta_{nm}\,,\\
\label{eq:continuant}
&&(E_i-a_n)g_n(E_i)=g_{n+1}(E_i)+b_n^2 g_{n-1}(E_i)\,.
\eea
Note that, by definition,
$g_{-1}(E_j)=0, g_{0}(E_j)=1$.
Furthermore, using the definition of wave function in terms
of the Krylov basis \eqref{phi} one can find the function 
$\phi_n$ in terms of $g_n(E_i)$ as
\be\label{phis}
\phi_n(t)=\frac{1}{b_n!}\sum_{j=1}^{{\cal D}_\psi}|c_j|^2 g_{n}(E_j)\, e^{-iE_j t}\,.
\ee
It is straightforward to see  that by plugging this expression into equation
\eqref{Sch-phi} one arrives  at \eqref{eq:continuant}.


\section{Krylov complexity probes chaos}
\label{sec:Krylov-chaos}

Endowed with the general relations derived in the previous section, in the present section, we establish the main claim of our paper, namely, we explore the key role of Krylov complexity in detecting the chaotic behaviour of a given system at a late time. 

Krylov complexity is defined as the expectation value of the number operator in the Krylov basis \cite{Parker:2018yvk}
\be\label{KC-def}
{\cal C}(t)=\langle \psi(t)|{\cal N}|\psi(t)\rangle\,,
\ee
where the number operator in the Krylov basis is defined as
\be
\label{eq:def-complexity}
{\cal N}=
\sum_{n=0}^{{\cal D}_\psi-1} \,n|n\rangle\langle n|\,.
\ee
To investigate the time dependence of complexity, we must therefore examine the time evolution of the expectation value \eqref{eq:def-complexity}. At the end of this section, we will be able to demonstrate that this expectation value, at late times, will be an important marker to detect the chaotic nature of any given system.

To proceed, let us consider, once again, the initial state 
\eqref{eq:psi0-Dpsi}. 
With this, the complexity \eqref{KC-def} reads
\be\label{KC}
{\cal C}(t)={\rm Tr}(\rho_{\rm DE}{\cal N})
+\sum_{j\neq k}^{{\cal D}_\psi}
e^{i\omega_{jk}t}c^*_j c_k {\cal N}_{jk} \,,
\ee
where $\omega_{jk}=E_j-E_k$, ${\cal N}_{jk}=\langle E_j|{\cal N}|E_k\rangle$ and the  diagonal density matrix
$\rho_{\rm DE}$ given by
\be
\label{eq:rho-DE}
\rho_{\rm DE}=
\sum_{j=1}^{{\cal{D}}_\psi} |c_j|^2
|E_j\rangle\langle E_j|\,.
\ee 
It is worth emphasizing once more, starting from the initial state \eqref{eq:psi0-Dpsi}, we only focus on a fixed charge sector of the full Hilbert space spanned by the Krylov basis. Accordingly, the diagonal density matrix $\rho_{\rm DE}$ defined in \eqref{eq:rho-DE} is confined to that particular symmetry-resolved block only.

The infinite time average of complexity is then given by
\be
\overline{ {\cal C}}=\lim_{T\rightarrow\infty}\frac{1}{T}\int_0^T
\langle {\cal N}(t)\rangle\, dt={\rm Tr}(\rho_{\rm DE}{\cal N})\,.
\ee

Knowing the explicit form of the number operator allows us to extract generic features of Krylov complexity. For this, it is  useful to write an explicit form of 
the matrix elements of the number operator. Indeed,
from equation \eqref{nE} one finds
\be
{\cal N}_{jk}=\sum_{n=0}^{{\cal D}_\psi-1}
n f_{nj} f^*_{nk}\,.
\ee
Thereafter, using \eqref{fnj1}, one arrives at
\be
\label{eq:N-cg}
{\cal N}_{jk}=c_j c^*_k\sum_{n=0}^{{\cal D}_\psi-1}
\frac{ng_n(E_j)g_n(E_k)}{\left(b_n!\right)^2}
\,.
\ee
Furthermore, by definition, the complexity is zero at $t = 0$,
so that, from \eqref{KC} we have
\be
{\cal C}(0)=\sum_{j, k=1}^{{\cal D}_\psi}
c^*_j c_k{\cal N}_{jk}=0\,.
\ee
This leads to 
\be\label{Nij}
\overline{{\cal C}}=\sum_{j}^{{\cal D}_\psi}
|c_j|^2 {\cal N}_{jj}=-
\sum_{j\neq k}^{{\cal D}_\psi}c^*_j c_k
{\cal N}_{jk}>0\,.
\ee

For finite many-body systems where the Lanczos coefficients vanish at the step $n = {\cal D}_\psi$, the Krylov complexity saturates\footnote{It is worth noting that for models where $b_n$ vanishes at $n = {\cal D}_\psi$, one may have a symmetry so that $b_n = b_{{\cal D}_\psi-n}$. In this case, the vanishing of the Lanczos coefficients does not result in complexity saturation, instead, we get a periodic behaviour, which could be due to an underlying $SU(2)$ symmetry \cite{Caputa:2021sib}. Indeed, the exact solution of equation \eqref{eq:continuant} based on Krawtchouk polynomials belongs to this category. Of course, in what follows, we will consider cases without such a symmetry, so that the complexity saturates at late times. }. Therefore, at late times, we have
\be\label{Otime}
{\cal C}(t)\approx {\cal C}_{\rm S}+
{\rm small\; fluctuations}\,,
\ee
where \( {\cal C}_{\rm S} \) is the saturation value of complexity.  This constant should be contrasted with the infinite time average
of complexity $\overline{{\cal C}}$. Actually, 
to achieve such a behaviour, the time-dependent phase in equation \eqref{KC} must sum to zero at late times, leading to a constant value equal to the late-time saturation value of the complexity. Indeed, from  
\eqref{KC} one finds
\be
\label{eqn:CbarC}
\overline{|{\cal C}(t)-\overline{{\cal C}}|^2}
=\sum_{j\neq k}^{{\cal D}_\psi}
|c_j|^2 |c_k|^2 {\cal N}^2_{jk} \,,
\ee
which demonstrates that 
$\overline{{\cal C}}\approx {\cal C}_{\rm S}$ iff 
the right-hand side of \eqref{eqn:CbarC} is 
sufficiently small. In the context of quantum thermalization, this is always  the case for 
operators that obey the ETH ansatz, where the off-diagonal matrix elements are exponentially small. This also ensures a rapid thermalization of any typical  ETH observable right after the system has relaxed.
On the contrary, it is important to note that the number operator is not an ETH operator. Consequently, it is not immediately clear whether the saturation value corresponds to the infinite time average
of complexity.

From \eqref{Nij} it is clear that the off-diagonal matrix elements of 
the number operator are of the same order as the 
diagonal ones. Therefore, in generic many-body systems with exponentially dense energy levels, the expression from \eqref{KC} suggests that if complexity saturates, the onset of saturation should occur at a time that is exponentially large in terms of the size of the system. This in turn makes the non-ETH-like behaviour of the number operator manifest.

In conclusion, we observe that from equation \eqref{KC}, for generic many-body systems, the Krylov complexity begins at zero value and continues to grow 
until it reaches an exponentially large time, at which point it saturates at a value which is of the order of the infinite time average of Krylov complexity. {\it Notably, this behaviour is a general characteristic of Krylov complexity in all many-body systems, regardless of whether they are integrable or chaotic.}

From equation \eqref{KC} and following the observation that complexity saturation begins at exponentially large times, we can actually estimate the saturation value of complexity. 
To proceed, we use the explicit form of the matrix elements of the number operator to rewrite the complexity \eqref{KC} as follows\footnote{To arrive at this result, we have used the fact that the expression for complexity obtained by plugging \eqref{eq:N-cg} into \eqref{KC}, is manifestly symmetric under the exchange of $j$ and $k$.}
\be\label{KCg}
{\cal C}(t)=\sum_{n=0}^{{\cal D}_\psi-1}\frac{-2n}{\left(b_n!\right)^2}\sum_{j\neq k}^{{\cal D}_\psi}
\sin^2(\frac{\omega_{jk}}{2}t)|c_j|^2 |c_k|^2
g_n(E_j)g_n(E_k) \,.
\ee

From this expression, it is evident that the complexity is zero at $t = 0$ as expected. Furthermore, it also manifests the expected quadratic growth at early times. Finally, 
it saturates at late times. To determine the saturation value, in what follows we examine the 
behaviour of complexity at large times.

Since we are interested in the complexity at the large time limit, it is useful to arrange the energy eigenvalues in increasing order, $E_1<E_2<\cdots<
E_{{\cal D}_\psi}$, and reformulate the summation in equation \eqref{KCg} as follows\footnote{ We note that Eq.~\eqref{eq:C-energyordered} naturally reproduces the expected behavior for energy eigenstates. In this case, the Krylov state expansion contains only one nonzero coefficient, say $c_{j_0} = 1$, with all others zero. Since the complexity expression involves sums over products $|c_{j+\ell}|^2 |c_j|^2$ with $\ell \geq 1$, each term in the sum vanishes, yielding $
{\cal C}(t) = 0$.
This aligns perfectly with the physical intuition that an energy eigenstate remains static in Krylov space and hence has zero Krylov complexity.}
\bea
\label{eq:C-energyordered}
&&{\cal C}(t)=-2\sum_{n=0}^{{\cal D}_\psi-1}\frac{n}{\left(b_n!\right)^2}\sum_{\ell=1}^{{\cal D}_\psi-1}
\sum_{j=1}^{{\cal D}_\psi-\ell}
\sin^2(\frac{s_{j+\ell}}{2}t)
\\ &&\hspace{3.1cm}\times\,|c_{j+\ell}|^2 |c_j|^2 g_n(E_{j+\ell})g_n(E_j) \,,\nonumber
\eea
where $s_{j+\ell}=E_{j+\ell}-E_j$. It is clear that the 
large-time behaviour of complexity depends on the 
distribution of $s_{j+\ell}$, which, in turn, relies on 
the distribution of energy eigenvalues. {\it This is where the information about the
	nature of the system, whether it is chaotic or integrable, 
	enters into the computations}.

Since the eigenvalues are arranged in an increasing order, 
the smallest values of $s_{j+\ell}$, namely the ones with $\ell=1$ oscillate slower than those with 
higher $\ell$ in the late time limit. 
The contributions coming from these slow oscillating modes towards the late-time behaviour of complexity are thus determined by the level spacing of the system. It is well known that for integrable systems, the level spacing follows a Poisson distribution, while for chaotic systems, it conforms to the Wigner surmise\footnote{
In general, for chaotic systems, the level spacing distribution follows the Wigner-Dyson form $
P(s) = \mu_0 s^\delta e^{-\mu_1 s^2}$,  where $\delta = 1, 2, 4$ corresponds to the Gaussian Orthogonal Ensemble (GOE), Gaussian Unitary Ensemble (GUE), and Gaussian Symplectic Ensemble (GSE), respectively. The constants $\mu_0$ and $\mu_1$ are ${\cal O}(1)$ normalization factors determined by the conditions 
$\int_0^\infty P(s)\,ds = 1,\;\int_0^\infty s P(s)\,ds = 1
$, which ensure that $P(s)$ is a properly normalized probability distribution with unit mean spacing.}
\be
\label{eq:distributions}
P_{\rm In}(s)=e^{-s},\;\;\;\;\;\;
P_{\rm Ch}(s)=\frac{\pi}{2}\,s\,e^{-\frac{\pi}{4}s^2}\,.
\ee

In chaotic systems, the distribution of $s_{j+1}$ exhibits\footnote{It is worth noting, however, $s_{j+l}$ appearing in \eqref{eq:C-energyordered} has the dimension of energy while the level spacing $s$ appearing in \eqref{eq:distributions} is dimensionless. Of 
	course, it is known that to find the universal level spacing
	an appropriate normalization is required \cite{Evnin:2018jbh}. } a peak around a value which is of order
one, and this behaviour extends to other $\ell$ as well. The $\sin^2$ factor oscillates between zero and one, and for large $t$, its average value approaches 
$\frac{1}{2}$. Thus, we can estimate the complexity ${\cal C}(t)$ at late times as
\be
{\cal C}(t)\approx -\sum_{n=0}^{{\cal D}_\psi-1}\frac{n}{\left(b_n!\right)^2}\sum_{j\neq k}^{{\cal D}_\psi}
|c_j|^2 |c_k|^2 g_n(E_j)g_n(E_k) \,.
\ee
This leads to
\be
\label{eq:C-latetime-chaotic}
{\cal C}(t)\approx \overline{{\cal C}}+
{\rm fluctuations}\,,
\ee
at late times. In order to identify the late time saturation value of complexity with the infinite time average, as in \eqref{eq:C-latetime-chaotic},  we have used \eqref{Nij}. 
The nature of these subleading fluctuations, however, depends on the 
specific state for which the complexity is calculated. 
It is worth emphasizing that the information about the late time saturation of complexity is encoded in the Lanczos coefficients 
appearing in the definition of $g_n(E_j)$.

On the other hand, for integrable models, the Poisson distribution suggests that the peak of level spacing for $s_{j+1}$ is centred around zero. Nevertheless, for large values of $\ell$ this peak shifts towards an ${\cal O}(1)$ value. Consequently, we need to decompose the complexity into two parts: the first part involves fast oscillations from $s_{j+\ell}$ 
for $\ell > 1$, and the second part, corresponding to $\ell = 1$. The latter represents a slowly oscillating term.

For large $t$, we can approximate $\sin^2$ by 
$\frac{1}{2}$ in the first part, leading to a constant contribution. However, since the first term does not consider $\ell = 1$, the resultant constant is smaller than the infinite time average of the complexity. Denoting this value as ${\cal C}_0 < \overline{{\cal C}}$, we find
\bea
&&{\cal C}(t)\approx {\cal C}_0-2\sum_{n=0}^{{\cal D}_\psi-1}\frac{n}{\left(b_n!\right)^2}
\sum_{j=1}^{{\cal D}_\psi-1}
\sin^2(\frac{s_{j+1}}{2}t)
\\ &&\hspace{3.1cm}\times\,|c_{j+1}|^2 |c_j|^2
g_n(E_{j+1})g_n(E_j) \,.\nonumber
\eea
For sufficiently large $t$, we can still estimate the 
$\sin^2$ of the second term as $\frac{1}{2}$, which results in complexity saturation approaching the infinite time average. However, for intermediate (yet still large) $t$, the second part oscillates, {\it indicating that for integrable models, complexity saturation remains below the infinite time average and approaches it from below as time progresses}.

Therefore, the saturation value of the Krylov 
complexity and the way it saturates in finite many-body systems could serve as a measure to determine whether the model is integrable or chaotic. To be precise, we observe that at the saturation time
\begin{empheq}[box=\fbox]{align}
	&{\cal C}_S\approx {\rm Tr}(\rho_{\rm DE}{\cal N}),\;\;\;
	\;\;\;\;\;\;{\rm for\;\;chaotic\;\; systems}\,,\\
	&{\cal C}_S < {\rm Tr}(\rho_{\rm DE}{\cal N}),\;\;\;\;
	\;\;\;\;\;{\rm for\;\; integrable\;\; systems}\,.
	\nonumber
\end{empheq}

It is also insightful to compute the time derivative of the complexity around the saturation time, which yields the dominant contribution of the form 
\be\label{firstD}
\dot{{\cal C}}(t)\approx
\sum_{j=1}^{{\cal D}_\psi-1}
\sin(s_{j+1}t)\,s_{j+1}
|c_{j+1}|^2 |c_j|^2
A(E_j,E_{j+1}) \,,
\ee
where $A(E_j,E_{j+1})>0$ is
\be
A(E_j,E_{j+1})=-\sum_{n=0}^{{\cal D}_\psi-1}\frac{n}{\left(b_n!\right)^2} g_n(E_{j+1})g_n(E_j) \,.
\ee

For integrable models, the expression \eqref{firstD} approaches 
zero due to the $s_\alpha$ factor. In 
contrast, for chaotic models, due to the level spacing 
distribution and for certain initial states this could 
in general result in a finite value, which 
may be almost zero at a particular time $t_m$ identified with the saturation time. Moreover, it is also
possible to show that the second derivative of 
this term is negative at that point\footnote{Note that for small but nonzero $s_{j+1}$—as is typical in chaotic systems—the extremum $t_m$ occurs near the point where $s_{j+1} t_m \sim \pi$. At this time, the derivative is negative because the $\cos(s_{j+1} t_m)$ factor becomes negative, while all other terms in Eq.~\eqref{firstD} remain positive. This implies a negative slope and supports the existence of a local maximum at $t = t_m$.
}, indicating 
that if such an extremum exists, it must be a maximum. 
Consequently, in this case, the complexity exhibits a 
peak before saturating at the value determined by the 
infinite time average of the complexity. Of course,
it is important to note that, in general, depending on the initial state, the peak may not be present at all or even if it is present, it may not be pronounced enough. 

Thus, a distinguishing signature of whether a model is chaotic or integrable can be signalled through how the Krylov complexity approaches its infinite time average at late times. {\it For integrable models, the complexity approaches the average from below, whereas for chaotic models, it approaches the saturation value at a finite time. Furthermore, depending on the initial state, it may also exhibit a peak before saturation}. 

We note, however, that for certain initial states of a chaotic system, the complexity does exhibit a clear peak before saturation.  To elaborate on this point further, we note that for any initial state $|\psi_0\rangle$ in a typical many-body system, one can define the canonical density matrix as follows
\be\label{Can-rho}
\rho_{\rm th}=\frac{e^{-\beta H}}{Z(\beta)},\;\;\;\;
Z(\beta)={\rm Tr}\left(e^{-\beta H}\right)\,,
\ee
where the inverse temperature $\beta$ is determined by 
the equation ${\rm Tr}(\rho_{\rm th} 
{ H })=\langle\psi_0|H|\psi_0\rangle=E_0$. 
It is also useful to define the {\it thermal Krlyov 
	complexity} as 
\be
{\cal C}_{\rm th}={\rm Tr}(\rho_{\rm th} {\cal N})\,,
\ee
which could be the saturation value of Krylov 
complexity for the initial state $|\psi_0\rangle$ if 
the number operator were an ETH operator. 

For chaotic systems, one would expect that expectation values of typical operators saturate to their thermal values shortly after 
relaxation, though as we have argued before, we would expect complexity to saturate at exponentially long times. Therefore, the saturation value of Krylov complexity should exceed the thermal 
Krylov complexity, which is attained at the thermalization time. This is a simple consequence of the fact that the thermalization time is significantly shorter than the time required for the 
complexity to saturate.

With these definitions in mind, let us now consider the following initial state
\be\label{TS}
|\psi_0\rangle=\frac{1}{\sqrt{Z(\beta)}}
\sum_{j=1}^{{\cal D}_\psi} e^{-\frac{1}{2}\beta E_j}
|E_j\rangle\,.
\ee
For this state, interestingly, the infinite time average 
of complexity equals the thermal Krylov complexity
\be
{\cal C}(t)={\cal C}_{\rm th}+
\frac{1}{Z(\beta)}\sum_{j\neq k}^{{\cal D}_\psi}
e^{-\frac{1}{2}\beta ( E_j+E_k)}
e^{i(E_j-E_k)t}{\cal N}_{jk}\,.
\ee

As mentioned above, for chaotic systems, the 
complexity exceeds the thermal Krylov complexity 
around the thermalization time and subsequently 
saturates to its infinite time average at longer 
times. For the initial state \eqref{TS}, the time 
average coincides with the thermal Krylov complexity, 
leading to the observation that the complexity 
exhibits a peak around the saturation time, 
approaching the infinite time average from above. In 
contrast, for integrable models, the complexity nearly 
reaches the thermal Krylov complexity and approaches 
its infinite time average from below at late times.

Thus, in a typical chaotic system the complexity
can indeed exhibit a clear peak before it saturates to the value given by its infinite time average, for certain initial states, the state given in \eqref{TS} being one such designated initial state.


\section{Explicit examples}\label{example}

In this section, we put our proposal to rigorous numerical tests using a particular many-body system, namely, the spin-$\frac{1}{2}$ Ising model described by the Hamiltonian
\be\label{Ising}
H=-J\sum_{i=1}^{N-1}\sigma^z_i\sigma^z_{i+1}-\sum_{i=1}^N
(g \sigma^x_i+h\sigma^z_i)\,.
\ee
Here $\sigma^{z,y,z}$ are Pauli matrices
and $J, g$ and $h$ are constants which define the model. By rescaling one can set $J=1$. Then the nature of the model, namely whether it is chaotic or integrable, is solely determined by the 
value of the constants $g$ and $h$. In particular,  
for $gh\neq 0$ the model is non-integrable. In order to perform our numerical computations we will set  $h=0.5,\, g=-1.05$ for chaotic case  \cite{Banuls:2011vuw}, and $h=0,\, g=-1.05$ to extract behaviour in the integrable regime. 

After fixing the above-mentioned values of the constants $J, g$ and $h$,  we numerically compute the Krylov complexity in this model for several initial states. We specifically consider three different initial states corresponding to configurations where all 
spins are aligned along the $x,y$ and $z$ 
directions, denoted as $|X+\rangle, |Y+\rangle,|Z+\rangle$, respectively. 

In the chaotic limit of this Ising model 
\eqref{Ising}, these states have been considered 
in \cite{Banuls:2011vuw} to explore the nature of thermalization, in particular, whether the thermalization is strong or weak\footnote{By making 
	use of certain measures the nature
	of thermalization for different spin models 
	has been studied in \cite{Sun:2020ybj,Lin_2017,Chen:2021,
		Bhattacharjee:2022qjw,Nandy:2023brt,prazeres2024continuoustransitionweakstrong,Alishahiha:2024rwm,Menzler:2024atb}.}. 
In the strong thermalization, the expectation  value 
of a typical operator relaxes to the thermal value very fast, while  for weak thermalization  it strongly oscillates around the thermal value, though its
time average attains the thermal value. It has been shown that although the initial state $|Y+\rangle$ exhibits strong thermalization, the initial states $|Z+\rangle$  and $|X+\rangle$ do not (see also \cite{Sun:2020ybj}). 

It is known that the Hamiltonian \eqref{Ising} has a parity symmetry which is essentially the reflection symmetry about the centre of the chain. Therefore, the energy eigenstates can be organized into two
blocks according to their parity, whether it is positive or negative. For $N=11$ which is the number of spin we will consider in our numerical computations, the dimension of subspace with the positive party is 1056 and that with negative parity is 992. 

It is easy to check that the aforementioned initial states  have positive parity which in turn indicates that the obtained Krylov subspace in chaotic case  should be a subspace with positive parity. One can compute the dimension of Krylov space associated with these states to obtain ${\cal D}_\psi=1056$. It is, however, important to note that for the integrable cases, besides the parity, due to extra possible conserved charges, the dimensions of Krylov spaces are actually smaller than this. In fact, one 
finds ${\cal D}_{Y,Z}=925$ and ${\cal D}_{X}=463$.

\begin{figure}[h!]
\begin{center}
	\includegraphics[width=0.49\linewidth]{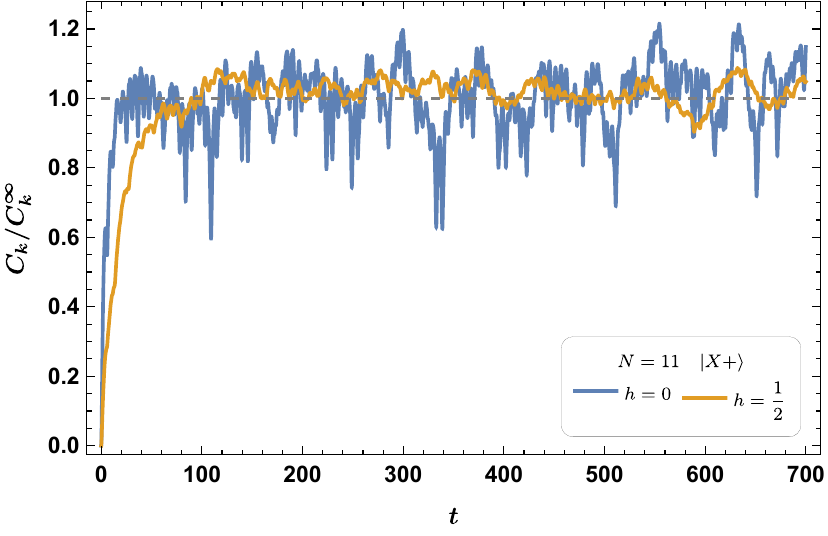}
	\includegraphics[width=0.49\linewidth]{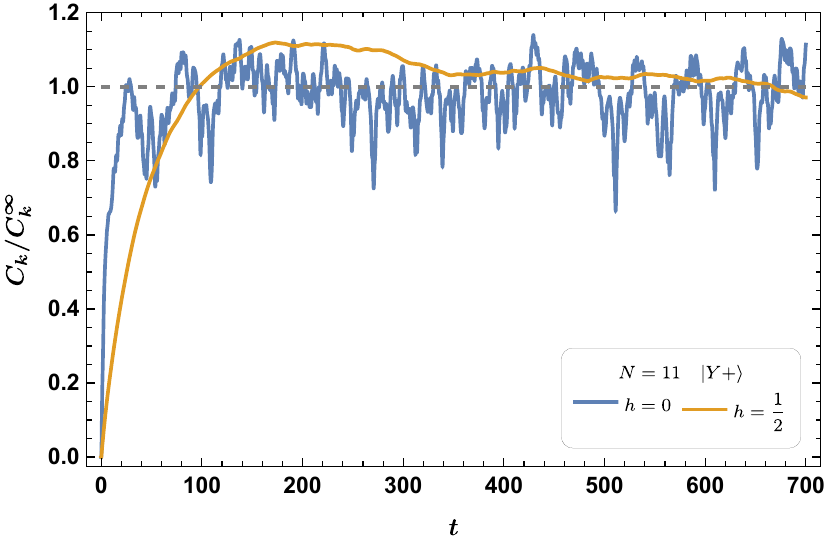}
	\includegraphics[width=0.49\linewidth]{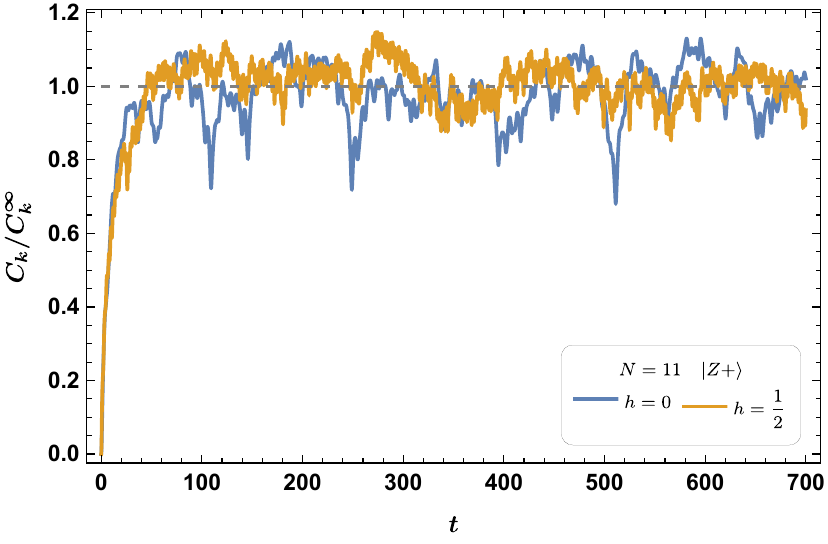}
\end{center}
\caption{Time evolution of complexity  for different initial states  $|X+\rangle, |Y+\rangle$   and $|Z+\rangle$ for the case where the model is integrable ($h=0$) and chaotic ($h=0.5$).  The computation is performed for $N=11$.
}\label{fig:complexity.intandchi1}
\end{figure}

It is straightforward to compute the Lanczos coefficients 
for these initial states from which one can compute complexity numerically\footnote{
	Lanczos coefficients for the model under consideration have also
	been computed in\cite{Noh_2021,Espa_ol_2023,Ballar_Trigueros_2022,Scialchi:2023bmw,Bhattacharya:2023zqt,Bhattacharya:2022gbz}
	.} by using \eqref{phis}. The numerical results for the
chaotic ($h=0.5$) and the integrable ($h=0$) cases are  depicted in 
figures \ref{fig:complexity.intandchi1}\footnote{
	We normalized the complexity by its infinite-time average value $(C_k^{\infty})$ to better illustrate how complexity approaches this average in both integrable and chaotic cases.}.

We observe that, although the numerical results are qualitatively consistent with the analytical argument presented in the previous section, different initial states can lead to significant fluctuations, making it difficult to clearly distinguish between chaotic and integrable behaviors. In particular, for the chaotic case and the initial state $|Y+\rangle$, the complexity shows a distinct peak before saturation, similar to that of the thermofield double state.

The magnitude and nature of these fluctuations depend on the initial states. Although the significant fluctuations around the saturation value do not affect the overall physical behavior of complexity, it is interesting to focus on initial states with milder fluctuations to observe clearer behavior, such as the $|Y+\rangle$ state in chaotic cases.


To identify an initial state with minimal fluctuations, we note that the inverse participation ratio (IPR), defined by

\be
\lambda^{-1} = \sum_{\alpha=1}^{\mathcal{D}} |c_\alpha|^4,
\;\;\;\;\;{\rm with}\;1\leq \lambda\leq {\mathcal{D}}\,,
\ee
may serve as a measure of the magnitude of fluctuations around the saturation value of complexity. Indeed, we have observed that  the larger the IPR, the milder the fluctuations in complexity.

More precisely, when the IPR value is sufficiently large (for our model, when $\lambda > 0.15\,\mathcal{D}$), fluctuations in complexity are suppressed, allowing for a clearer observation of its expected behavior in both chaotic and integrable states.

In Figure 2, we consider a special initial state where the IPR is large enough in both integrable and chaotic cases, ensuring that fluctuations do not obscure the complexity dynamics. The corresponding initial state is
\be\label{ISYZ}
|\psi_0\rangle=|Y+\rangle_1\otimes |Z-\rangle_2\otimes
|Y+\rangle_3\otimes |Z-\rangle_4\otimes\cdots\,.
\ee
\begin{figure}[h!]
\begin{center}
	\includegraphics[width=0.65\linewidth]{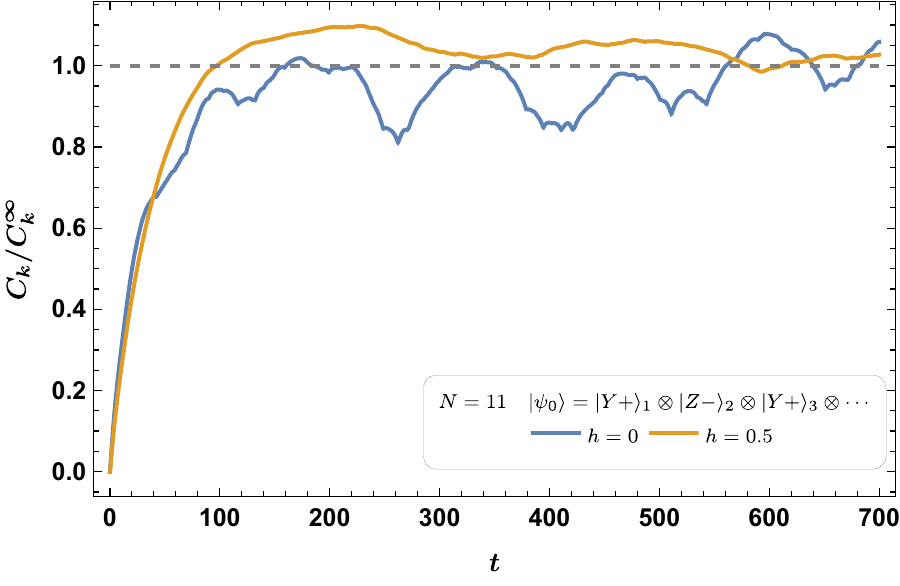}
\end{center}
\caption{Time evolution of complexity  for  initial state given by \eqref{ISYZ} for the case where the model is integrable ($h=0$) and chaotic ($h=0.5$).  The computation is performed for $N=11$.
}\label{fig:complexity.intandchi2}
\end{figure}

\begin{figure}[h!]
\begin{center}
	\includegraphics[width=0.65\linewidth]{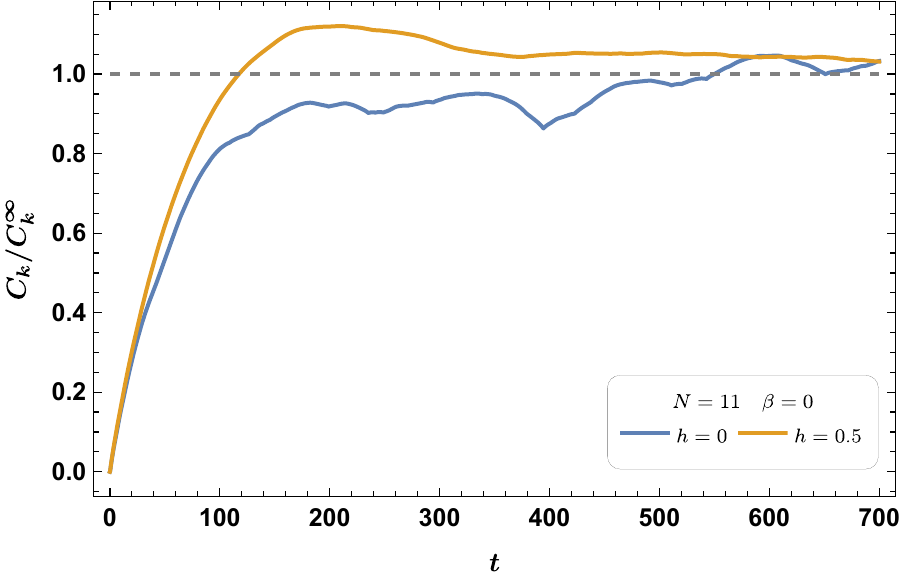}
\end{center}
\caption{Time evolution of complexity  for initial state  given by \eqref{TS}
	with $\beta=0$. One observes that in the chaotic case the complexity exhibits 
	a clear peak before saturation. The computation is performed for $N=11$.
}\label{fig:complexity.TS}
\end{figure}

In order to crosscheck our analytic claim regarding the TFD state, 
we have also computed complexity for 
the initial state \eqref{TS} for $\beta=0$\footnote{It is worth noting that this initial state has maximal IPR (i.e., equal to ${\cal D}$), making it a very suitable choice for studying the behavior of complexity saturation.}. 
The result is shown in figure \eqref{fig:complexity.TS}.
From this figure one can immediately observe that the complexity for chaotic case does exhibit a peak before saturation. 
Furthermore, the behavior of complexity for the integrable case also confirms the other part of our claim, namely it approaches the 
infinite time average from below and reaches it 
at late times.

These plots show distinctive behavior of complexity for chaotic and integrable models. While the complexity saturates to its
saturation value given by the infinite time average of complexity at the saturation time for chaotic, in the integrable case it reaches this value from below at large times.

\begin{figure}[h!]
\begin{center}
	\includegraphics[width=0.49\linewidth]{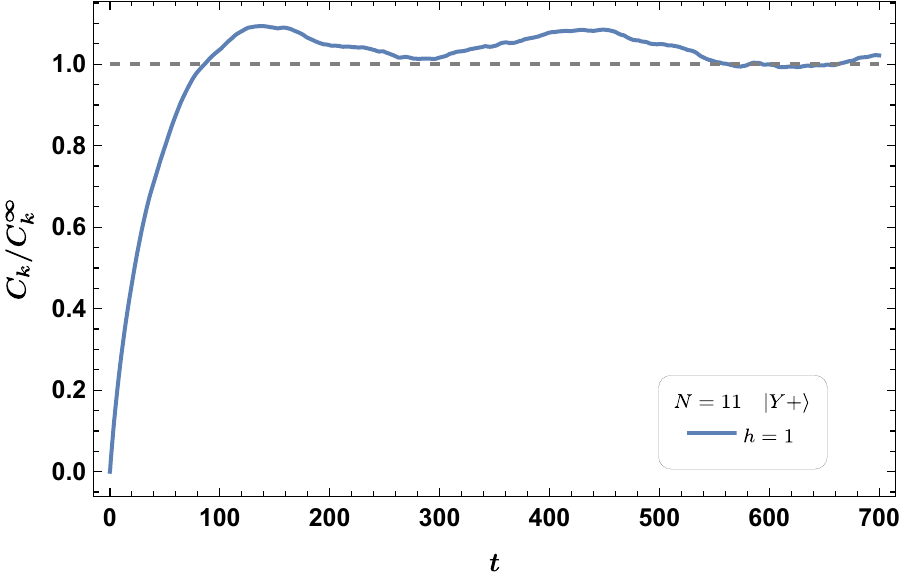}
	\includegraphics[width=0.49\linewidth]{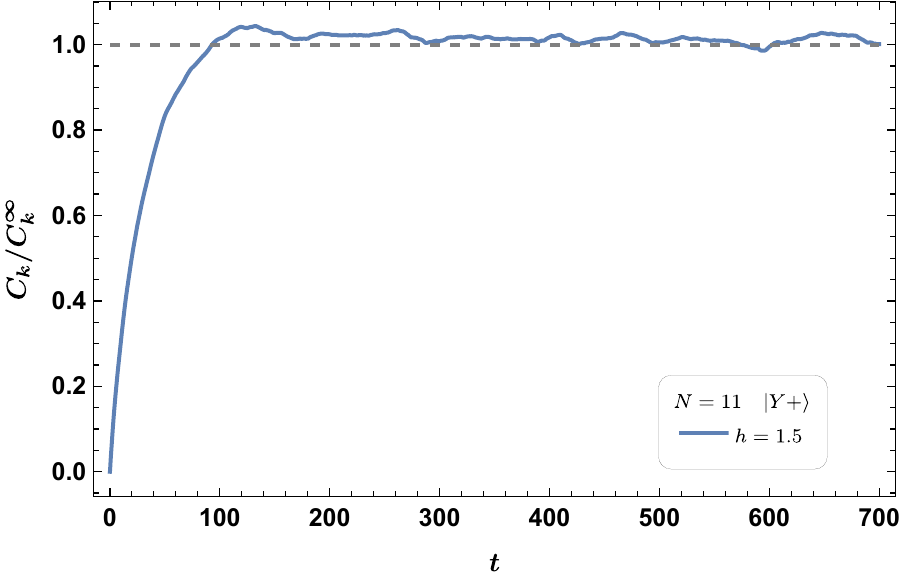}
	\includegraphics[width=0.49\linewidth]{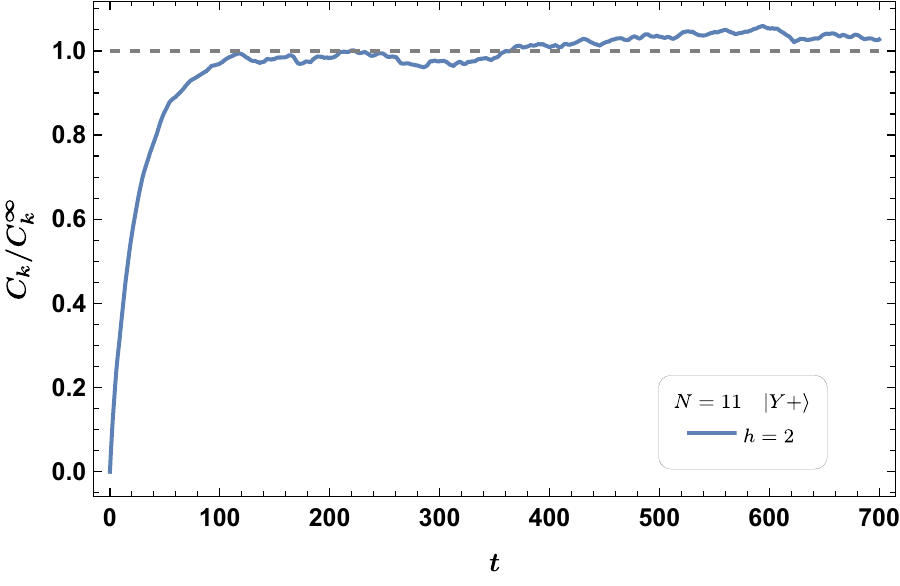}
	\includegraphics[width=0.49\linewidth]{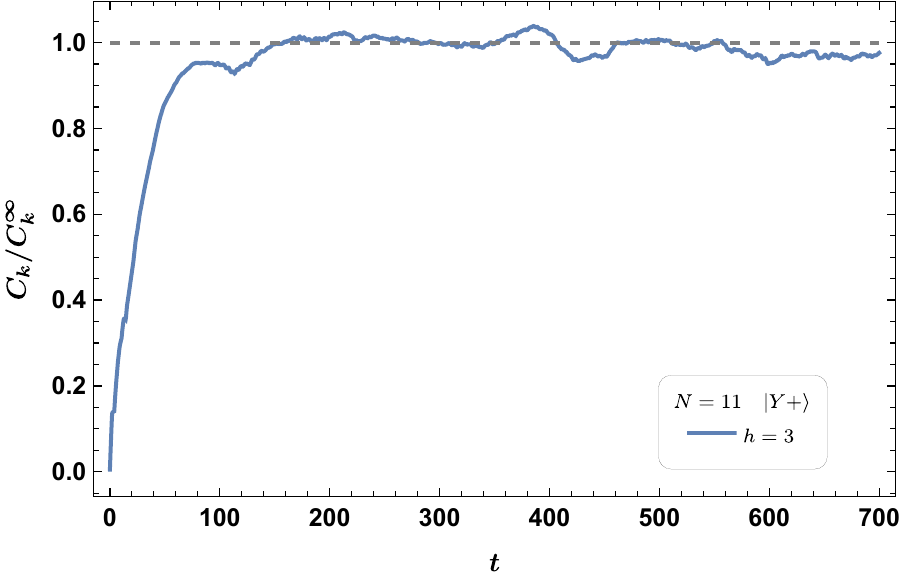}
\end{center}
\caption{Time evolution of complexity  for initial state $|Y+\rangle$.
	for $h=1,1.5,2, 3$. This shows that as we progress
	towards higher values of $h$, the system moves away from
	chaos and leans towards integrability. The computation is performed for $N=11$.
}\label{fig:complexity.h}
\end{figure}

Thus far, our numerical computations have been centred on two specific 
points within the parameter space of the model \eqref{Ising}, 
specifically $(g=-1.05, h=0)$ and $(g=-1.05, h=0.5)$. These points 
correspond to integrable and nearly maximally chaotic scenarios, leading 
to Poisson and Wigner distributions for level spacing, respectively. It would be more interesting to explore additional points and calculate the Krylov complexity, which would allow us to determine whether the system tends towards chaotic or integrable behaviours depending on the parametric closeness of the system to the aforementioned points. In pursuit of this, we have also evaluated the 
Krylov complexity for the initial state $|Y+\rangle$ at different values for $h$, yielding the results depicted in figure \eqref{fig:complexity.h}. 
Analysis of these figures reveals that as we progress towards higher 
values of $h$, the system moves away from chaos and leans towards 
integrability, as expected. This observation further supports our earlier 
conclusion in the preceding section that the $\sin$-term and level 
spacing play crucial roles in determining the behaviour of Krylov complexity at late times.

\begin{figure}[h!]
\begin{center}
	\includegraphics[width=0.49\linewidth]{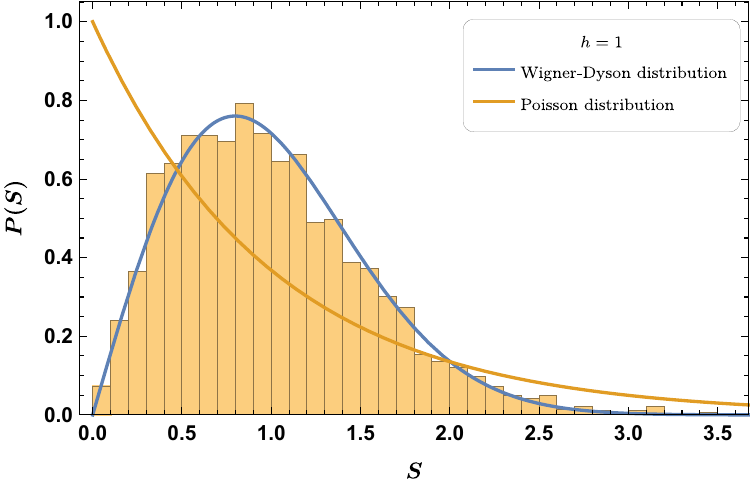}
	\includegraphics[width=0.49\linewidth]{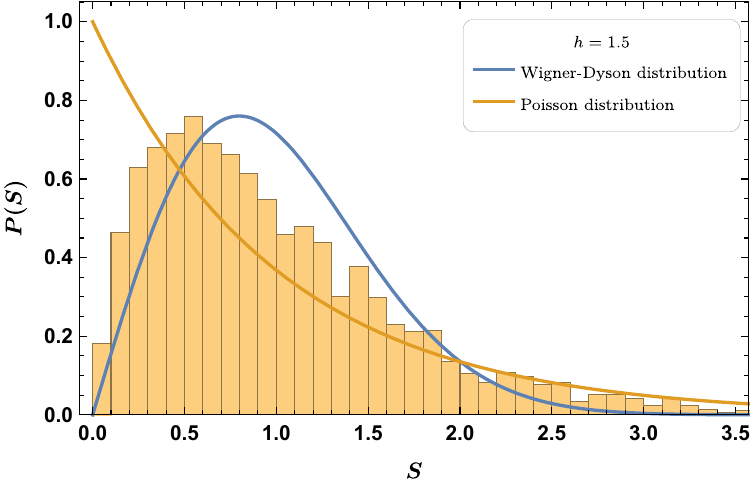}
	\includegraphics[width=0.49\linewidth]{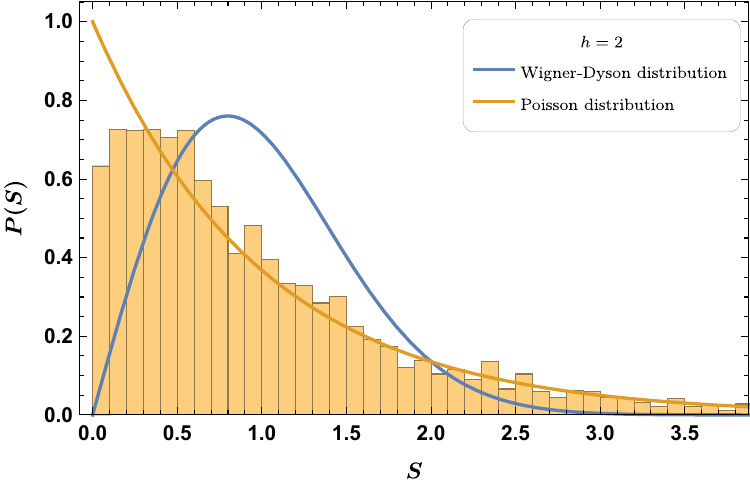}
	\includegraphics[width=0.49\linewidth]{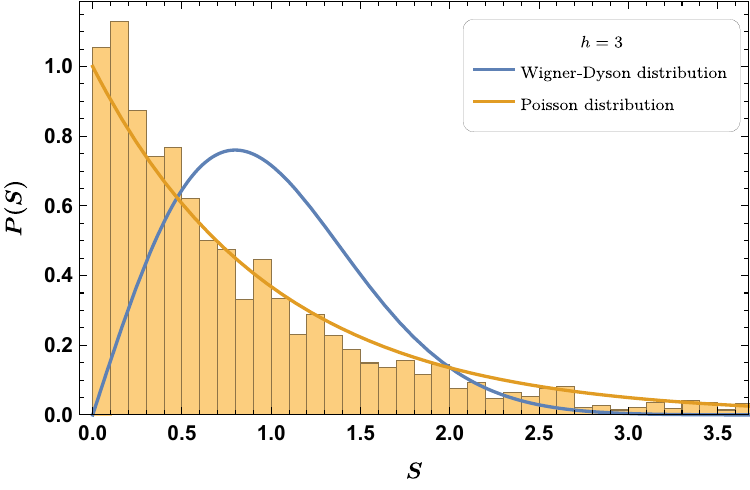}
\end{center}
\caption{Level spacing for energy eigenvalues of the model
	\eqref{Ising} with positive parity for different values of $h$ given by $h=1,1.5,2, 3$. 
	This shows that as we progress
	towards higher values of $h$, the distribution moves away from
	Wigner  and leans towards Poisson. To have a better statistic the computation is performed for $N=13$.
}\label{fig:level}
\end{figure}

Furthermore, it is instructive to examine the level spacing for different values of $h$ which we present in Figure \ref{fig:complexity.h}. The corresponding level spacings are shown in Figure \ref{fig:level}. The latter clearly demonstrates how the level spacing transitions from a Wigner distribution to a Poisson distribution as $h$ increases.


\section{Discussions}

In this work, we have explored whether Krylov complexity can 
effectively probe the chaotic nature of many-body quantum 
systems. To address this outstanding question, we have utilized the fact that Krylov complexity is defined as the expectation value of the number operator in the Krylov basis. Inspired by the quantum 
thermalization process, we have examined the time evolution of 
complexity, observing that it exhibits some general common features which are not specific to the nature of the particular system in consideration. 
In particular, we show that at exponentially long times, the Krylov complexity saturates (asymptotically) at a value very close to the infinite time average of complexity, regardless of whether the system is chaotic or integrable.

Nevertheless, we demonstrate that the manner in which complexity saturates does depend, significantly, on the dynamical nature of the 
systems. For a generic initial state in a typical chaotic system, the 
saturation value of complexity reaches its infinite time average 
at the saturation time, whereas in integrable models, 
it approaches the infinite time average value from 
below over a long duration.

Indeed, we established a connection between complexity and level spacing statistics, a well-known measure of chaos. This relationship enabled us to explicitly demonstrate why and how complexity can serve as an effective indicator of chaos.

However, it is important to note that the exact value of 
complexity at the saturation point and its detailed behaviour at large and intermediate times depend not only on the dynamics but also on the initial state used for computing the Krylov complexity. Specifically, when the initial state is a maximally entangled state, the Krylov complexity, while approaching the infinite time average from above, leads to a peak before ultimately saturating at that average value. The appearance of this peak is not universal or state-independent. In fact, for a generic initial state, the complexity may or may not exhibit a clear peak. To clarify, for chaotic systems, the infinite-time average of complexity surpasses the thermal Krylov complexity. Consequently, the expression ${\rm Tr} \left((\rho_{{\rm DE}}-\rho_{{\rm th}}){\cal N}\right)$ is non-negative. Therefore, we observe a more pronounced peak for those states where this expression is near zero.

Moreover, there are small fluctuations around the saturation value whose shapes also depend on the initial state. This confirms the postulates of \cite{Susskind:2014moa,Brown:2016wib} for complexity saturation. We also found that these fluctuations are directly related to the inverse participation ratio (IPR). A state with a sufficiently large IPR exhibits the expected behavior more clearly and reliably.

Our observation regarding the behaviour of Krylov complexity is 
founded on the explicit form of the number operator within the 
Krylov basis. The Krylov complexity can be expressed in the 
following general form:
\be
\mathcal{C}(t) = \sum_{j \neq k}^{\mathcal{D}\psi} 
\sin^2\left(\frac{\omega_{jk}}{2} t\right) |c_j|^2 |c_k|^2 A(E_j, E_k),
\ee
where $A(E_j, E_k)$
is a smooth function of its arguments which can be read by comparing 
this equation with \eqref{KCg}. It 
is evident that the late-time behaviour of complexity is 
dictated by the distribution of $\omega_{jk} = E_j - E_k$, a 
quantity that may provide insights into the chaotic behaviour 
of the system. Importantly, one must ensure that only those 
energy eigenvalues that lie within a symmetric block are 
included. Notably, the Lanczos algorithm used to construct the 
Krylov basis inherently satisfies this constraint.

It is important to highlight that achieving the desired saturation behaviour requires the function $A(E_j, E_k)$ to be smooth and not develop poles, a condition that holds for finite $n$. In infinite-dimensional systems, where $n$ can become very large, we can analyze the late-time behaviour of complexity by adopting the continuum limit. This approach enables us to compute the function 
$A(E_j, E_k)$, which often reveals a double-pole structure, resulting in linear growth of complexity at late times \cite{Alishahiha:2022nhe, Alishahiha:2022anw}.

Finally, it is worth noting that for a thermal initial state where 
$|c_j|^2=e^{-\beta E_j}/Z[\beta]$, the Krylov complexity can be expressed in a form which is reminiscent of the spectral complexity introduced in \cite{Iliesiu:2021ari}
(see also \cite{Camargo:2023eev}). Therefore, we would expect that the spectral complexity exhibits the same late-time behaviour as that of the Krylov complexity.

\section*{acknowledgements}

We would like to thank Rathindra Nath Das, Johanna Erdmenger, René Meyer, 
Reza Pirmoradian, Mohammad Reza Tanhayi and Hamed Zolfi for useful 
discussions. This work is based upon research founded by Iran National 
Science Foundation (INSF) under project No 4023620.


\begin{thebibliography}{widestlabel}

\bibitem{Bohigas:1983er}Bohigas, O., Giannoni, M. \& Schmit, C. Characterization of chaotic quantum spectra and universality of level fluctuation laws. {\em Phys. Rev. Lett.}. \textbf{52} pp. 1-4 (1984)
\bibitem{Shenker:2013yza}Shenker, S. \& Stanford, D. Multiple Shocks. {\em JHEP}. \textbf{12} pp. 046 (2014)
\bibitem{Shenker:2013pqa}Shenker, S. \& Stanford, D. Black holes and the butterfly effect. {\em JHEP}. \textbf{3} pp. 067 (2014)
\bibitem{Maldacena:2015waa}Maldacena, J., Shenker, S. \& Stanford, D. A bound on chaos. {\em JHEP}. \textbf{8} pp. 106 (2016)
\bibitem{Kitaev}Kitaev, A. A Simple Model of Quantum Holography. 
\bibitem{Sachdev:1992fk}Sachdev, S. \& Ye, J. Gapless spin fluid ground state in a random, quantum Heisenberg magnet. {\em Phys. Rev. Lett.}. \textbf{70} pp. 3339 (1993)
\bibitem{Maldacena:2016hyu}Maldacena, J. \& Stanford, D. Remarks on the Sachdev-Ye-Kitaev model. {\em Phys. Rev. D}. \textbf{94}, 106002 (2016)
\bibitem{Fine_2014}Fine, B., Elsayed, T., Kropf, C. \& Wijn, A. Absence of exponential sensitivity to small perturbations in nonintegrable systems of spins 1/2. {\em Physical Review E}. \textbf{89} (2014,1)
\bibitem{Xu_2019}Xu, S. \& Swingle, B. Accessing scrambling using matrix product operators. {\em Nature Physics}. \textbf{16}, 199-204 (2019,11)
\bibitem{Hashimoto:2020xfr}Hashimoto, K., Huh, K., Kim, K. \& Watanabe, R. Exponential growth of out-of-time-order correlator without chaos: inverted harmonic oscillator. {\em JHEP}. \textbf{11} pp. 068 (2020)
\bibitem{Balasubramanian:2022tpr}Balasubramanian, V., Caputa, P., Magan, J. \& Wu, Q. Quantum chaos and the complexity of spread of states. {\em Phys. Rev. D}. \textbf{106}, 046007 (2022)
\bibitem{Alishahiha:2022anw}Alishahiha, M. \& Banerjee, S. A universal approach to Krylov state and operator complexities. {\em SciPost Phys.}. \textbf{15}, 080 (2023)	
\bibitem{Caputa:2024vrn}Caputa, P., Jeong, H., Liu, S., Pedraza, J. \& Qu, L. Krylov complexity of density matrix operators. {\em JHEP}. \textbf{5} pp. 337 (2024)
\bibitem{viswanath1994recursion}Viswanath, V. \& Müller, G. The Recursion Method: Application to Many Body Dynamics. (Springer Berlin Heidelberg,1994)
\bibitem{Parker:2018yvk}Parker, D., Cao, X., Avdoshkin, A., Scaffidi, T. \& Altman, E. A Universal Operator Growth Hypothesis. {\em Phys. Rev. X}. \textbf{9}, 041017 (2019)
\bibitem{Nandy:2024htc}Nandy, P., Matsoukas-Roubeas, A., Martinez-Azcona, P., Dymarsky, A. \& Campo, A. Quantum Dynamics in Krylov Space: Methods and Applications.  (2024,5)
\bibitem{Chu:2024cah}Chu, Y., Li, X. \& Cai, J. Quantum delocalization on correlation landscape: The key to exponentially fast multipartite entanglement generation.  (2024,4)
\bibitem{Dymarsky:2021bjq}Dymarsky, A. \& Smolkin, M. Krylov complexity in conformal field theory. {\em Phys. Rev. D}. \textbf{104}, L081702 (2021)
\bibitem{Bhattacharjee_2022}Bhattacharjee, B., Cao, X., Nandy, P. \& Pathak, T. Krylov complexity in saddle-dominated scrambling. {\em Journal Of High Energy Physics}. \textbf{2022} (2022,5)
\bibitem{Avdoshkin:2022xuw}Avdoshkin, A., Dymarsky, A. \& Smolkin, M. Krylov complexity in quantum field theory, and beyond. {\em JHEP}. \textbf{6} pp. 066 (2024)
\bibitem{Camargo:2022rnt}
H.~A.~Camargo, V.~Jahnke, K.~Y.~Kim and M.~Nishida,
JHEP \textbf{05}, 226 (2023)
doi:10.1007/JHEP05(2023)226
[arXiv:2212.14702 [hep-th]].
\bibitem{Vasli:2023syq}Vasli, M., Babaei Velni, K., Mohammadi Mozaffar, M., Mollabashi, A. \& Alishahiha, M. Krylov complexity in Lifshitz-type scalar field theories. {\em Eur. Phys. J. C}. \textbf{84}, 235 (2024)
\bibitem{Rabinovici:2021qqt}Rabinovici, E., Sánchez-Garrido, A., Shir, R. \& Sonner, J. Krylov localization and suppression of complexity. {\em JHEP}. \textbf{3} pp. 211 (2022)
\bibitem{Rabinovici:2022beu}Rabinovici, E., Sánchez-Garrido, A., Shir, R. \& Sonner, J. Krylov complexity from integrability to chaos. {\em JHEP}. \textbf{7} pp. 151 (2022)
\bibitem{Scialchi:2023bmw}Scialchi, G., Roncaglia, A. \& Wisniacki, D. Integrability-to-chaos transition through the Krylov approach for state evolution. {\em Phys. Rev. E}. \textbf{109}, 054209 (2024)
\bibitem{Ballar_Trigueros_2022}Ballar Trigueros, F. \& Lin, C. Krylov complexity of many-body localization: Operator localization in Krylov basis. {\em SciPost Physics}. \textbf{13} (2022,8)
\bibitem{Espa_ol_2023}Español, B. \& Wisniacki, D. Assessing the saturation of Krylov complexity as a measure of chaos. {\em Physical Review E}. \textbf{107} (2023,2)
\bibitem{Erdmenger:2023wjg}Erdmenger, J.,Jian, S. \& Xian, Z. Universal chaotic dynamics from Krylov space. {\em JHEP}. \textbf{8} pp. 176 (2023)
\bibitem{Huh:2023jxt}Huh, K., Jeong, H. \& , J. J. F. Pedraza, Spread complexity in saddle-dominated scrambling. {\em JHEP}. \textbf{5} pp. 137 (2024)
\bibitem{Camargo:2024deu}Camargo, H., Huh, K., Jahnke, V., Jeong, H., Kim, K. \& Nishida, M. Spread and Spectral Complexity in Quantum Spin Chains: from Integrability to Chaos.  (2024,5)
\bibitem{Nandy:2024wwv}Nandy, P., Pathak, T. \& Tezuka, M. Probing quantum chaos through singular-value correlations in sparse non-Hermitian SYK model.  (2024,6)
\bibitem{Bhattacharjee:2024yxj}Bhattacharjee, B. \& Nandy, P. Krylov fractality and complexity in generic random matrix ensembles.  (2024,7)
\bibitem{Balasubramanian:2024ghv}Balasubramanian, V., Das, R., Erdmenger, J. \& Xian, Z. Chaos and integrability in triangular billiards.  (2024,7)
\bibitem{Baggioli:2024wbz}Baggioli, M., Huh, K., Jeong, H., Kim, K. \& Pedraza, J. Krylov complexity as an order parameter for quantum chaotic-integrable transitions.  (2024,7)
\bibitem{Alishahiha:2022nhe}Alishahiha, M. On quantum complexity. {\em Phys. Lett. B}. \textbf{842} pp. 137979 (2023)
\bibitem{Srednicki:1994mfb}Srednicki, M. Chaos and Quantum Thermalization. {\em Phys. Rev. E}. \textbf{50} (1994,3)
\bibitem{Deutsch:1991}Deutsch, J. Quantum statistical mechanics in a closed system. {\em Phys. Rev. A}. \textbf{43}, 2046-2049 (1991,2)
\bibitem{Lanczos:1950zz}Lanczos, C. An iteration method for the solution of the eigenvalue problem of linear differential and integral operators. {\em J. Res. Natl. Bur. Stand. B}. \textbf{45} pp. 255-282 (1950)
\bibitem{PhysRevB.103.134207}Herviou, L., Bardarson, J. \& Regnault, N. Many-body localization in a fragmented Hilbert space. {\em Phys. Rev. B}. \textbf{103}, 134207 (2021,4)
\bibitem{Balasubramanian:2022dnj}Balasubramanian, V., Magan, J. \& Wu, Q. Tridiagonalizing random matrices. {\em Phys. Rev. D}. \textbf{107}, 126001 (2023)
\bibitem{miki2022singleindexedexceptionalkrawtchoukpolynomials}Miki, H., Tsujimoto, S. \& Vinet, L. The single-indexed exceptional Krawtchouk polynomials.  (2022)
\bibitem{Caputa:2021sib}Caputa, P., Magan, J. \& Patramanis, D. Geometry of Krylov complexity. {\em Phys. Rev. Res.}. \textbf{4}, 013041 (2022)
\bibitem{Evnin:2018jbh}Evnin, O. \& Piensuk, W. Quantum resonant systems, integrable and chaotic. {\em J. Phys. A}. \textbf{52}, 025102 (2019)
\bibitem{Banuls:2011vuw}Bañuls, M., Cirac, J. \& Hastings, M. Strong and Weak Thermalization of Infinite Nonintegrable Quantum Systems. {\em Phys. Rev. Lett.}. \textbf{106}, 050405 (2011)
\bibitem{Sun:2020ybj}Sun, Z., Cui, J. \& Fan, H. Quantum information scrambling in the presence of weak and strong thermalization. {\em Phys. Rev. A}. \textbf{104}, 022405 (2021)
\bibitem{Lin_2017}Lin, C. \& Motrunich, O. Quasiparticle explanation of the weak-thermalization regime under quench in a nonintegrable quantum spin chain. {\em Physical Review A}. \textbf{95} (2017,2)
\bibitem{Chen:2021}Chen, F., Sun, Z., Gong, M., Zhu, Q., Zhang, Y., Wu, Y., Ye, Y., Zha, C., Li, S., Guo, S., Qian, H., Huang, H., Yu, J., Deng, H., Rong, H., Lin, J., Xu, Y., Sun, L., Guo, C., Li, N., Liang, F., Peng, C., Fan, H., Zhu, X. \& Pan, J. Observation of Strong and Weak Thermalization in a Superconducting Quantum Processor. {\em Phys. Rev. Lett.}. \textbf{127}, 020602 (2021,7)
\bibitem{Bhattacharjee:2022qjw}Bhattacharjee, B., Sur, S. \& Nandy, P. Probing quantum scars and weak ergodicity breaking through quantum complexity. {\em Phys. Rev. B}. \textbf{106}, 205150 (2022)
\bibitem{Nandy:2023brt}Nandy, S., Mukherjee, B., Bhattacharyya, A. \& Banerjee, A. Quantum state complexity meets many-body scars. {\em J. Phys. Condens. Matter}. \textbf{36}, 155601 (2024)
\bibitem{prazeres2024continuoustransitionweakstrong}Prazeres, L. \& Oliveira, T. Continuous Transition Between Weak and Strong Thermalization using Rigorous Bounds on Equilibration of Isolated Systems.  (2024)
\bibitem{Alishahiha:2024rwm}Alishahiha, M. \& Vasli, M. Thermalization in Krylov Basis.  (2024,3)
\bibitem{Menzler:2024atb}Menzler, H. \& Jha, R. Krylov localization as a probe for ergodicity breaking.  (2024,3)
\bibitem{Noh_2021}Noh, J. Operator growth in the transverse-field Ising spin chain with integrability-breaking longitudinal field. {\em Physical Review E}. \textbf{104} (2021,9)
\bibitem{Bhattacharya:2023zqt}Bhattacharya, A., Nandy, P., Nath, P. \& Sahu, H. On Krylov complexity in open systems: an approach via bi-Lanczos algorithm. {\em JHEP}. \textbf{12} pp. 066 (2023)
\bibitem{Bhattacharya:2022gbz}Bhattacharya, A., Nandy, P., Nath, P. \& Sahu, H. Operator growth and Krylov construction in dissipative open quantum systems. {\em JHEP}. \textbf{12} pp. 081 (2022)
\bibitem{Susskind:2014moa}Susskind, L. Entanglement is not enough. {\em Fortsch. Phys.}. \textbf{64} pp. 49-71 (2016)	
\bibitem{Brown:2016wib}Brown, A., Susskind, L. \& Zhao, Y. Quantum Complexity and Negative Curvature. {\em Phys. Rev. D}. \textbf{95}, 045010 (2017)
\bibitem{Iliesiu:2021ari}Iliesiu, L., Mezei, M. \& Sárosi, G. The volume of the black hole interior at late times. {\em JHEP}. \textbf{7} pp. 073 (2022)	
\bibitem{Camargo:2023eev}Camargo, H., Jahnke, V., Jeong, H., Kim, K. \& Nishida, M. Spectral and Krylov complexity in billiard systems. {\em Phys. Rev. D}. \textbf{109}, 046017 (2024)	
















\end{thebibliography}

\end{document}